\documentstyle[preprint,aps]{revtex}

\input epsf
\preprint{\bf PREPRINT \today}

\begin{document}
\columnsep0.1truecm
\draft
\title{The phase plane of moving discrete breathers}
\author{Paul A. Houle}
\address{Laboratory of Atomic and Solid State Physics, Cornell University, Ithaca, NY, 14853-2501}
\maketitle

\begin{abstract}
We study anharmonic localization in a periodic five atom chain with
quadratic-quartic spring potential. 
We use discrete symmetries to
eliminate the degeneracies of the harmonic chain and easily
find periodic orbits. 
We apply linear stability analysis to
measure the frequency of phonon-like disturbances in the
presence of breathers and to analyze the instabilities of
breathers.  We visualize the phase plane of 
breather motion directly and develop a technique for
exciting pinned and moving breathers.  We observe long-lived
breathers that move chaotically and a global transition to
chaos that prevents forming moving breathers at high energies.
\end{abstract}

\narrowtext

A body of theoretical work has appeared in the past decade
on intrinsic localized modes in perfect anharmonic
crystals. \cite{flachReview}
Although the existence of breather periodic orbits 
with localized energy is established \cite{POproof},  
less is known about the rest of the phase space.
In this paper,  we use a periodic chain of five atoms 
coupled by a quadratic-quartic springs
as a model for larger chains and as a testbed for
techniques which can be applied to other systems.
We use discrete reflection symmetries
to find submanifolds on which finding
breather solutions is simplified 
and on which breather solutions can be
continued to the phonon limit.
We introduce coordinates for visualizing
breather motion,  exhibiting the movability separatrix introduced in 
\cite{flachSeparatrix},
and detecting chaos in the separatrix region which leads to
long-lived breathers that move erratically.
We also apply linear stability analysis to investigate phonon-like
excitations in the presence of a breather and
to map the stable and unstable manifolds of an unstable breather,
developing a technique to systematically
launch moving and pinned breathers complementary to existing
methods. \cite{ChenMobility}.

\par

Introducing our system,  the Hamiltonian is

\begin{equation}
H = \sum_{i=1}^N M { p_i^2 \over 2 } +
\sum_{i=1}^N U(q_i - q_{i-1}),\ N=5\label{eq:hamiltonian}
\end{equation}

\begin{equation}
U(x) = { K_2 \over 2} x^2 + { K_4 \over 4} x^4 \label{eq:potential}.
\end{equation}

under condition that $K_4 > 0.$
By scaling $q_i$ and $p_i$ we set $M=K_2=K_4=1$ without loss of generality.
The system has four nontrivial degrees of freedom since center
of mass momentum is conserved.
To get numerical results we integrate the equations of motion 
for (\ref{eq:hamiltonian}) by fifth-order
Runge-Kutta with a fixed step size of $10^{-3}.$ \cite{NR}
Because a high amplitude breather is localized on a few atoms,  it
is reasonable that our small model
captures much of the physics of
larger chains.  To check this we transplanted both
even and odd breather solutions of the $N=5$ chain at energy $E=3$ 
onto a $N=20$ chain with all other atoms
initially at rest with zero displacement;  we  found 
the breather remained localized for over one thousand breather
oscillations
without obvious loss of amplitude.

\par

We take advantage of discrete symmetries to explore a subspace of
the complete phase space of the chain;  because breather periodic orbits
lie on these subspaces,  we can use symmetry to find breathers
and to follow breather solutions all the way to the $E=0$ phonon
limit.
The Hamiltonian (\ref{eq:hamiltonian}) respects two kinds of
reflection symmetry which we refer to as {\it even} and {\it odd} symmetry
after the even-parity and odd-parity 
breathers in prior literature.  \cite{StabMotion}
The displacements and momenta of an {\it even breather} have
even parity with respect to reflections around a bond.
(the pattern of displacements is
roughly ${\bf q} = (-{1 \over 6},1,-1,{1 \over 6},0)$ for a 
even breather on the bond between atoms 2 and 3).
We refer to this property as
{\it even symmetry} and the submanifold of the phase space with
even symmetry as the {\it even manifold.}  
The odd-parity periodic orbit  has {\it odd symmetry}, 
odd parity with respect to reflections around an atom.
(displacements are roughly ${\bf q}=(0,-{1 \over 2},1,-{1 \over 2},0)$
for an odd breather
centered on atom 3)  The submanifold of the chain with
odd symmetry is the {\it odd manifold.}
Even and odd symmetry are respected by the
dynamics;  if the system starts in the even or odd submanifold
it remains there for all time.
Even and odd manifolds,  of course,  exist for each site but as the
chain is translationally invariant the dynamics are identical at
all sites.
Even and odd symmetry can be used to simplify
any sized chain --- the use of symmetry is
particularly advantageous for the $N=5$ chain since the even
and odd submanifolds are reduced to two degree of freedom 
systems which are simple to understand and visualize.

\par
Unlike chains with an even number of atoms
\cite{flachBifurcate} \cite{sanduskyBifurcate} there is
no energy threshold for breathers in
a chain with an odd number of atoms.
As the energy $E \rightarrow 0$ we expect the anharmonic
chain to be approximated by a harmonic chain for
short times.  For N=5,
the harmonic chain has two degenerate pairs of normal
modes with wavenumbers $k_n={n \pi \over 5}$ and
frequencies $\omega_{\pm 1} = 2 \sin {\pi \over 5}$ and
$\omega_{\pm 2} = 2 \sin {2 \pi \over 5}$.
The modes with the highest $k$ are {\it band-edge} modes.
In general,  chains with an even number of atoms have a single
band edge mode with wavenumber $\pi$ and chains with an odd
number of atoms have two degenerate band edge modes.
Since any linear combination of two phonons with the
same frequency is a periodic orbit of a harmonic chain,
each degenerate pair of phonons is associated with a
two parameter family of periodic orbits.
This degenerate
situation is structurally unstable and is
shattered when the slightest amount of $K_4$ is turned on.
Since only one linear combination of phonons intersects each
submanifold,  we can remove the degeneracy by restricting ourselves
to an even or odd manifold on which a single parameter
family of periodic orbits survives in the anharmonic system.
Specifically,  a band-edge phonon with
even or odd symmetry deforms continuously into an even
or odd breather as energy increases.
No bifurcation occurs,  thus there is no energy
threshold for the formation of a breather in an $N$ odd chain.
The situation is different from that in
$N$ even chains for which there exists a single band-edge mode 
possessing both even and odd symmetry which 
undergoes a symmetry-breaking tangent bifurcation at an energy proportional
to $N^{-1}$ into two different periodic orbits corresponding
to both even and odd breathers.  
\cite{flachBifurcate} \cite{sanduskyBifurcate} 
\par

\begin{figure}
\centerline{\epsfxsize=7in\epsfbox{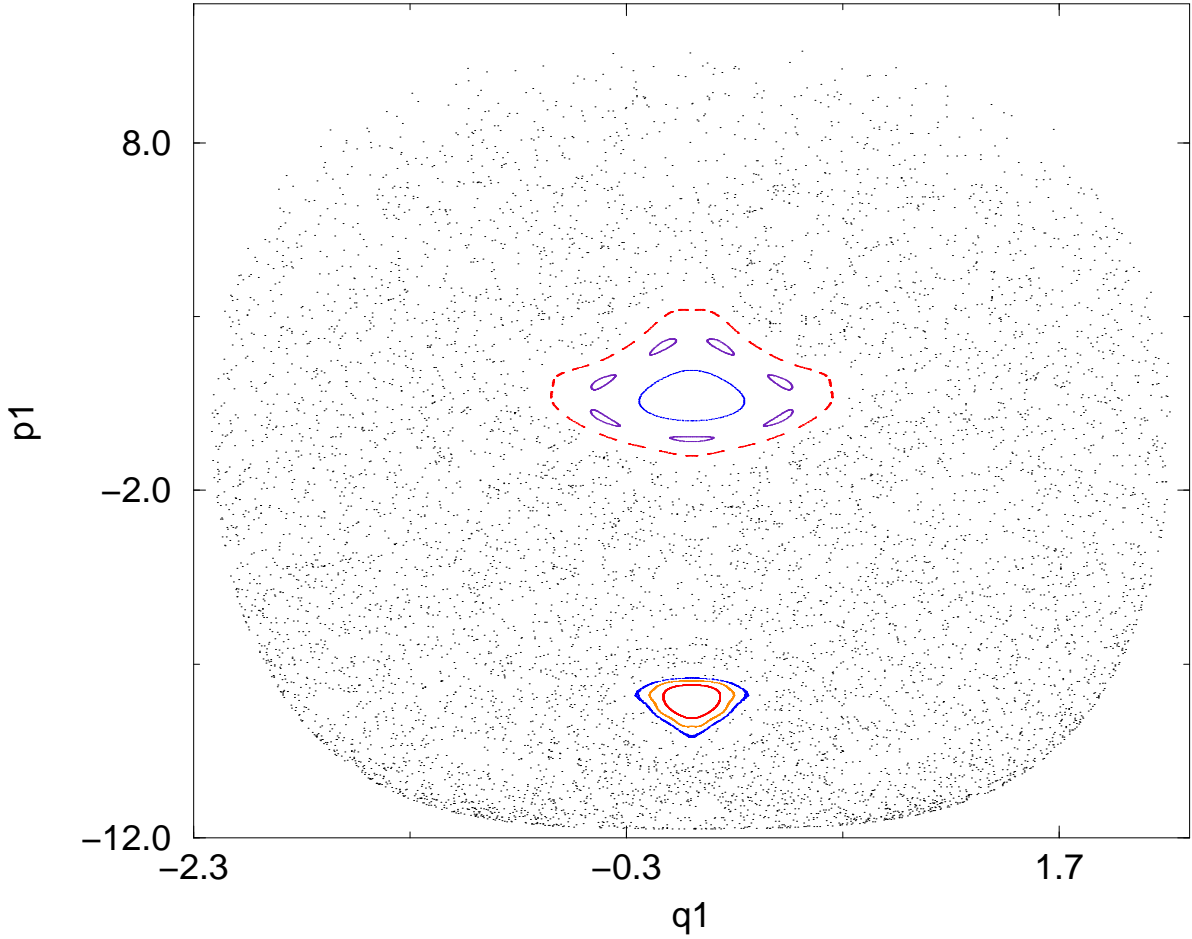}}
\caption{A Poinc\'{a}re section on the odd submanifold at $E=230;$
the phase space is dominated by chaotic trajectories,  although
two large islands of regularity are visible.  The upper island consists
of an unstable breather and phonon-like disturbances around it while
the lower island is due to a high-amplitude standing wave
associated with the $n=\pm 1$ phonons;  resonant islands in the upper
half of the plot involve phase locking between the breather and
quasi-phonons}
\label{fig:prettyPoincare}
\end{figure}

Poinc\'{a}re sections are effective for visualizing
dynamics on the submanifolds and provide a method
of finding breather solutions.  
Although restriction to submanifolds should simplify the
search for breathers for a system of any size,  it
is very advantageous for $N=5$ where the problem is
reduced to a one dimensional root search.
Periodic orbits manifest as fixed points of the Poinc\'{a}re map
on the surface of section $q_3 = 0, \dot{q}_3>0.$  We plot
surfaces of section by integrating the
equation of motion until the trajectory crosses the surface of
section --- at this point we use Newton-Raphson to solve
for the duration of a Runge-Kutta step that lands on the
surface of section.
Fixed points corresponding to
breather periodic orbits lie on the $q_1=0$ line and can
be found by a simple 1-d root
search by the Brent algorithm.  \cite{netlib}
Both the even and odd manifolds can be visualized by plotting
$(q_1,p_1)$;  since the submanifolds are two-dimensional,  the
result is a complete description of the dynamics on a submanifold.
Fig. \ref{fig:prettyPoincare} is an example.
Chaos is prevalent in odd manifold sections above $E=1$;
we have not observed obvious chaos in the even manifold in
the range $0<E<300$ that we've studied.

\par

\begin{figure}
\centerline{\epsfxsize=7in\epsfbox{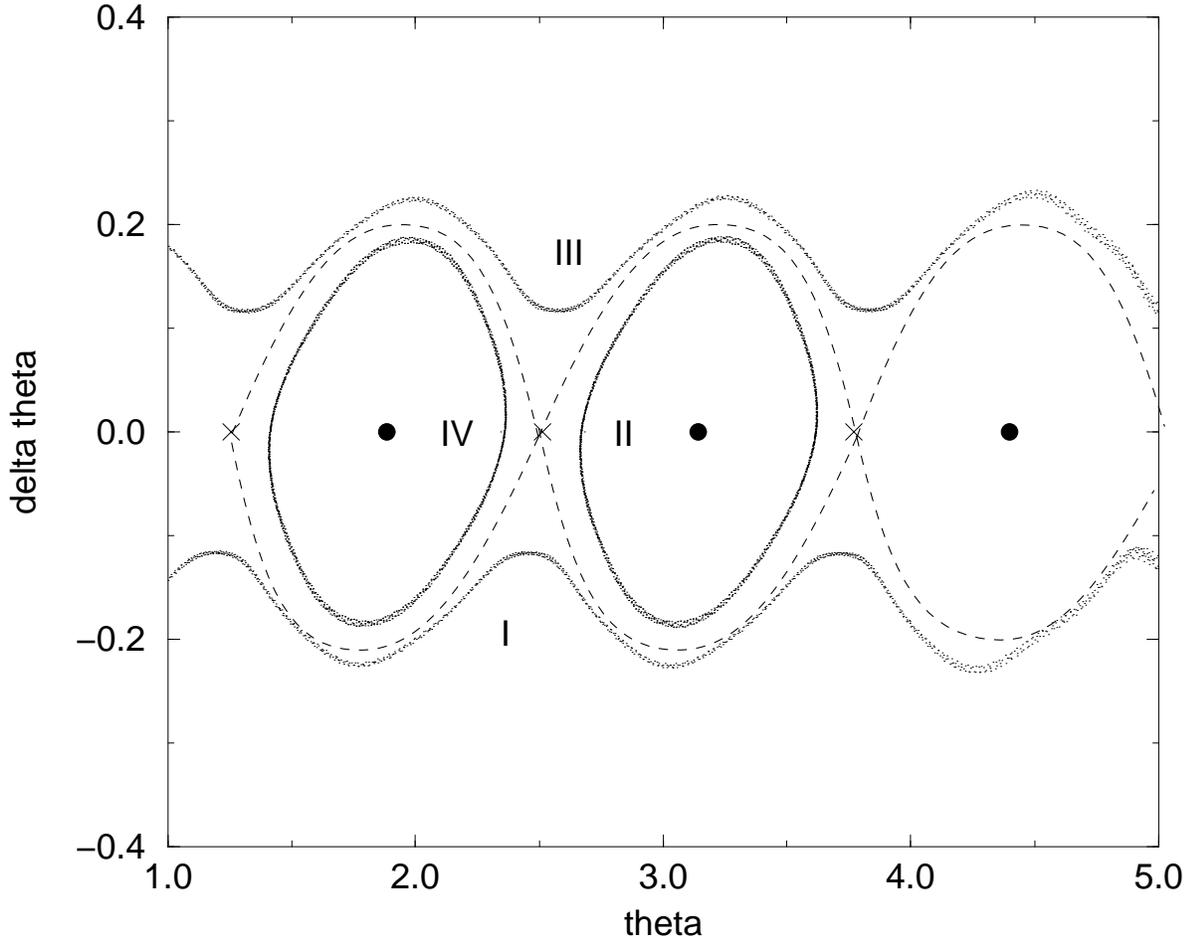}}
\caption{A Poinc\'{a}re section using the $q_3=0$ trigger and the
variable $\theta$ illustrating the pendulum-like phase plane
of moving breathers.  The dotted line is drawn in and illustrates
the separatrix;  circles mark stable
(even) breather solutions and crosses mark unstable
(odd) breather solutions.
Roman numerals indicate four regions in phase space divided
by stable and unstable manifolds in which distinct regular behaviors
are observed.  In regions I and III the breather moves to the left
and right respectively while it is pinned at either side of site
three in regions II and IV.}
\label{fig:motion}
\end{figure}

\begin{figure}
\centerline{\epsfxsize=7in\epsfbox{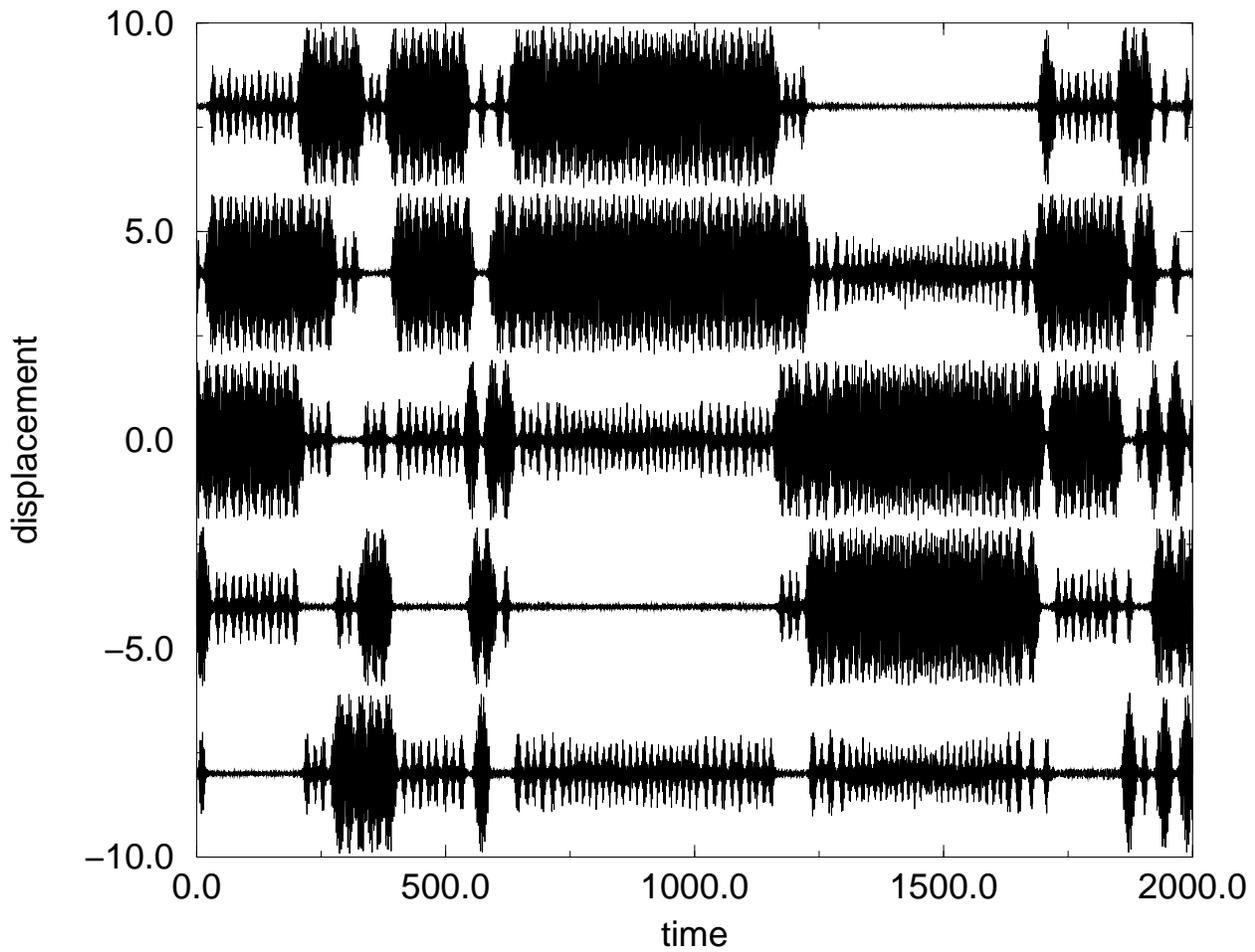}}
\caption{Hopping chaos observed at $E=50$ for a small perturbation in
region III.  Although energy is localized throughout the duration
of the simulation,  the breather moves erratically.
The atoms are visually separated by adding constants to the displacements;
the chains is periodic so the bottom atom is adjacent to the top atom.}
\label{fig:new_hopping}
\end{figure}

We use surfaces of section with a different method of projection
to directly visualize the movability
separatrix introduced in \cite{flachSeparatrix} in a manner
that should be useful for characterizing
moving and pinned modes in other breather systems.  
The phase plane of breather motion at
$E=10$ is seen in Fig. \ref{fig:motion}.
The location of the breather is given by the angle
$\theta = {\rm Arg}\ h$ where

\begin{equation}
h = \sum_{i=1}^N { p^2_i \over 2 } e^{i{2 \pi \over N} n}
+ \sum_i^N U(q_i-q_{i-1}) e^{i{2 \pi \over N}(n+{1 \over 2})},
\label{eq:hquantity}
\end{equation}

with N=5.  
(\ref{eq:hquantity}) is a complex weighted average of the
kinetic and potential energy on each atom and bond that
takes in account the circular nature of the chain,
a strong influence on small chains.  \cite{footnote1}  
We construct a variable conjugate to $\theta$ 
by treating $\delta \theta = \theta_n-\theta_{n-1}.$  
as the velocity of the breather.
The $q_3=0$ trigger works well when a breather is
localized in the range $1<\theta<5.$ \cite{footnote2}
Unlike Fig. \ref{fig:prettyPoincare},  Fig. \ref{fig:motion} is
a projection from a high-dimensional space to the plane and
is not a complete picture of the dynamics.
The curves in
Fig. \ref{fig:motion} is consistent with 
the hypothesis \cite{flachSeparatrix} \cite{flachOneReduced} that 
two sets of action-angle variables exist approximately for
breather states;  one set connected with the breather's spatial
position and velocity and a set of ``internal'' degree of freedoms
associated with breathing.  As occurs for the
pendulum,  the position-velocity conserved surfaces change topology at the
separatrix dividing pinned and moving states.  Even the
slightest degree of nonintegrability 
breaks conserved surfaces in the separatrix creating a
layer of chaos.  \cite{LandL}
We refer to the chaos observed in this separatrix 
as {\it hopping chaos}
because trajectories within the separatrix region 
move erratically in space while remaining localized for
thousands of oscillations or more as observed in Fig. \ref{fig:new_hopping}.
Hopping breathers have been observed to decay,  so it is clear that the
hopping chaos region of phase space is connected to delocalized chaotic
regions;  however,  the region of hopping chaos appears
to be sufficiently hemmed in by KAM tori that hopping chaos
is a distinct intermediate-term behavior.
The region of hopping chaos enlarges as energy increases;  near $E=20$
a global transition to chaos in the phase plane of
Fig \ref{fig:motion}
appears to occur and it becomes impossible
to create moving modes;  only the islands of near integrability
corresponding to pinning on a bond remain.
Hopping chaos is probably less robust in longer chains since
longer chains have more degrees of freedom for 
resonances to occur with and for
energy to be radiated into.

\par

Numerical linear stability analysis gives a local picture of the
phase space around a periodic orbit complementary to the more global
views of previous sections.  With stability analysis we
examine phonon-like excitations in the presence of breathers and
analyze the instabilities of breathers.
Let us write the state of the system
as a phase space vector ${\bf x} = ({\bf q},{\bf p})$;
let ${\bf x_0}$ be a point on a periodic orbit with period $T$.
We make an  infinitesimal change $\delta {\bf x}$ in the initial
conditions,  launching the system at time $t=0$ in
state ${\bf x}={\bf x}_0+\delta {\bf x}$.  When we observe the
system at time $t=T$ the system is in state 
${\bf x}_T = {\bf x}_0 + \delta {\bf x}_T.$  To linear
order in $\delta {\bf x}$ 

\begin{equation}
\delta {\bf x}_T =  S \delta {\bf x_0} + O(\delta {\bf x}^2_0)
\label{eq:stabMatrix}
\end{equation}

where $S$ is the {\it stability matrix.}

\par

With an accurately known periodic orbit we can determine $S$
numerically by making copies of the system and
perturbing them successively in each position and momentum
coordinate and then evolving each system for time $T.$
\cite{footnote3}
To interpret the stability matrix,  we first find eigenvalues and
eigenvectors with EISPACK.  \cite{netlib}
For a Hamiltonian system $S$ is a symplectic matrix with
certain constraints on the eigenvalues and eigenvectors.  \cite{LandL} 
In our application,  eigenvalues values come in three kinds of
pairs;  elliptic
pairs $\lambda=e^{\pm i \phi}$ indicating phonon-like
excitations in the presence of a breather,  hyperbolic pairs 
$\lambda_1=\lambda_2^{-1}$ with $\lambda_1$ real indicating
instabilities and
parabolic pairs $\lambda_1=\lambda_2=1$ arising from conserved
quantities.  In our application
two parabolic pairs arise due to
conserved quantities;  one pair
due to conservation of momentum and another due to
conservation of energy --- these uninteresting pairs
are removed by automated inspection of eigenvectors.

\par

\begin{figure}
\centerline{\epsfxsize=7in\epsfbox{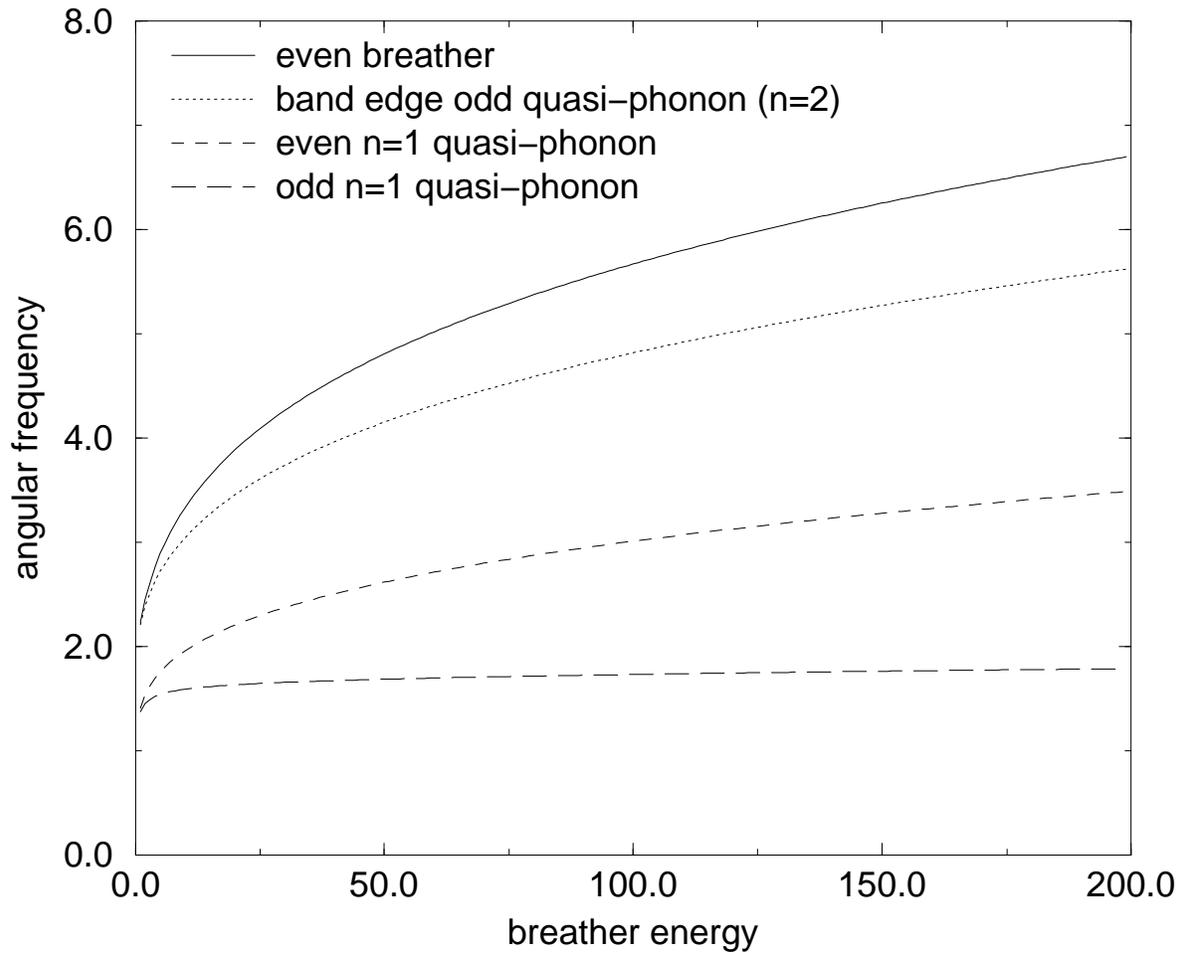}}
\caption{Breather and quasi-phonon frequencies from the stability
matrix for even breathers on a 5-atom quadratic-quartic chain.
The wavenumber of a quasi-phonon is $k=\pm {2 \pi \over 5}n$}
\label{fig:frequencies}
\end{figure}

Stability analysis confirms that the even breather known to be
stable in long chains \cite{StabMotion} is linearly stable in the
five atom chain;  no hyperbolic eigenvalues appear in the energy
range from $E=0$ -- $200$.  If a stable breather is infinitesimally
perturbed,  one excites phonon-like
disturbances that we call {\it quasi-phonons};  by investigating
quasi-phonons one can study the interaction between a breather
and phonons,  crucial for understanding quantum and thermal
fluctuations around breathers.  \cite{aubryQP}
We determine the frequencies of quasi-phonons from eigenvalues of the
stability matrix;  
Fig \ref{fig:frequencies} plots quasi-phonon frequencies as
a function of breather energy.  
To test the accuracy of our technique,  we excited 
quasi-phonons by making a small change in the
$E=3$ and $E=50$ breather solutions;  quasi-phonon frequencies appeared
as peaks in the power spectrum determined by running the
system until $t=3000$ and taking the FFT of one atom's position as a
function of time.
The two methods agree
to within one part in $10^{-4}$,  the bin size of
the power spectrum.  
Quasi-phonon frequencies vary smoothly with breather
energy and converge on the true phonon frequencies as the energy
of the breather goes to zero.  Therefore we label quasi-phonons
by the symmetry they uphold and the 
wavenumber $k_n$ of the phonon they become in the
$E \rightarrow 0$ limit,  a scheme that should remain applicable
for quasi-phonons in larger chains and higher dimensions.
The $n=2$ quasi-phonon is tangent to the phase plane of
breather motion;  a small excitation of the $n=2$ quasi-phonon
causes the breather to rock around one bond 
while a large excitation causes the breather to break free and move
as has been observed in a $\phi^4$ lattice. \cite{ChenMobility} 

\par

Perturbation of a stable breather has been used to create moving
breathers in a $\phi^4$ lattice \cite{ChenMobility};
we have found an alternate method of creating pinned and gliding breathers
by perturbing an unstable
breather.  As is known for large chains, \cite{StabMotion}
the odd breather is linearly unstable for all energies;  the
linear instability does not cause the breather to decay but
instead causes the breather to move.
The eigenvectors of $S$ with eigenvalues less than and greater
than one
point respectively into
the stable and unstable manifolds and
define a plane tangent to the phase plane of Fig. \ref{fig:motion}.
As the energy of the odd breather goes to zero,  the
unstable mode eigenvectors converge on the even band-edge phonon.
The tangent plane is divided into four quadrants by the stable
and unstable manifolds;  distinct regular behaviors are observed for
perturbations directed into each quadrant as illustrated
by the drawn-in separatrix in Figure \ref{fig:motion} ---
the unstable manifolds of an odd breather at one site feed into
the stable manifolds of odd breathers at neighboring sites.
By choosing a quadrant we can
launch a breather that travels either to the
left or to right or
that remains pinned while rocking slightly to the left or right of
the odd breather location.
To get reliable results it is necessary to add a sufficiently
large perturbation so as to clear the region of hopping 
chaos in the separatrix region.

\par

In summary,  we have found the $N=5$ chain exhibits much of the phenomenology
of larger chains and can be used as both a model of localization
and a testing ground for techniques.
We have applied numerical stability analysis to
to measure quasi-phonon
frequencies in the presence of a breather and to design perturbations
that create either moving or pinned modes starting from either a stable
or an unstable breather.  Chaos exists in the movability
separatrix and becomes increasingly prevalent as the
energy increases;  a global transition to chaos near
$E=20$ makes it impossible to launch moving breathers.
With appropriate coordinates we have been able to
directly visualize the phase plane of breather motion.

\par

I would like to thank C. Henley for suggesting 
equation (\ref{eq:hquantity}) and other useful discussions as
well as Rongji Lai,  S. A. Kiselev
and A. J. Sievers.  This work was supported by
by NSF Grant DMR-9612304.
\eject

\eject

\eject

\eject


\begin{references}
\bibitem{footnote1}{A similar variable applicable only for large chains
is used in \cite{flachSeparatrix}}
\bibitem{footnote2}{Triggering fails near $\theta=0$;  perfect triggering
is difficult since it is impossible to assign alternating signs to
five atoms in a ring.}
\bibitem{footnote3}{Our increment was $10^{-6}$,  the stability matrix
was insensitive to changes in the increment.}
\end{references}
\end{document}